\documentclass[aps,prd,superscriptaddress,nofootinbib,groupedaddress,amsfonts,floatfix,onecolumn]{revtex4}
\usepackage{graphicx,amsmath,amssymb,amstext,amssymb,amsbsy,amsfonts,amsthm,lscape,units,tabularx,multirow}

\newcommand{\beq}{\begin{equation}}
\newcommand{\eeq}{\end{equation}}

\usepackage{color}

\begin{document}

\title{Constraint structure  of the Generalized Proca model in  the Lagrangian formalism}

\newcommand{\IPM}{School of Astronomy, Institute for Research in Fundamental Sciences (IPM), P.O. Box 19395-5531, Tehran, Iran}

\author{Zahra Molaee}
\affiliation{\IPM}

\email{zmolaee@ipm.ir}
\author{Ahmad Shirzad}
\newcommand{\IUT}{Department of Physics, Isfahan University of Technology, Isfahan, Iran}
\affiliation{\IUT}

\begin{abstract}

We present a new Lagrangian approach for  the dynamical structure of the generalized Proca theory (GP). This approach includes the A-Z constraint structure of the model in the Lagrangian formalism and ends up with an accurate count of the number of degrees of freedom. We also give the complete Hamiltonian constraint structure of the model.  

\end{abstract}

\maketitle

\section{Introduction }

 Searching for the most general theory of  massive spin-1 particles is an interesting issue that has progressed much over the past decade. There have been some attempts to consider a theory with derivative self-interactions for the vector field\cite{a14}-\cite{a13}. Furthermore,  variety of the  extensions of these theories  and studying the corresponding  constraint algebra  can be found in  \cite{a21}-\cite{A18}.

 The most familiar method for studying the constrained systems concerns Hamiltonian formalism.
 It should be noted that  in the Proca theory, the primary  Hamiltonian constraint  is the consequence of the absence of time derivative of the temporal component of the vector field. 
Consistency  of this primary constraint generates a secondary constraint. The Poisson bracket of the secondary  and primary constraints  is proportional  to the mass of the field;  thus  the system is first class in the massless case and   second class in the massive case. 
 Massless spin-1 field theory describes  two propagating degrees of freedom (i.e. helicity $= \pm 1$); while, massive spin-1 field  theory has  three propagating physical degrees of freedom (i.e. helicity $= 0, \pm1$). 
 In other words, the mass term in Proca action breaks the $U(1)$ gauge invariance and invites the longitudinal polarisation to the theory.

  It is not difficult to see that generalized Proca theories (GP)  are not gauge theories, although they include a number of constraints. In fact, in the framework of Hamiltonian formalism, all of the constraints are second class. Analyzing the Hamiltonian dynamics of the system shows that all GP theories do have three dynamical degrees of freedom. These results  concerning the Faddeev-Jackiw  formalism  can be found in ref   \cite{c1,c2} . 
  In this paper we have given the Hamiltonian analysis of GP theories, in the frame work of Dirac approach (section 4), in order to have a complete study of the model. 
  
  However, our main objective is investigating the Lagrangian constraint structure of the model. Lagrangian approach for GP theories has been considered first in \cite{a14}. An important point in the Proca theory is that the component $A_0$ is non-dynamical, ensuring the absence of the Ostrogradsky ghost. In fact, one more constraint is needed to establish the correct number of the degrees of freedom . For GP theories it is somehow understood \cite{a14} that two constraints are needed to reach the correct number of degrees of freedom. This point was again emphasized in \cite{a18},  where it is argued that we can not insure about the total number of constraints just by finding the first one. Moreover, one should be convinced that the system is not a gauge theory. To this end, however, one needs to find exactly the complete set of Lagrangian constraints to be able to judge about the true number of degrees of freedom. 
  
  In this paper we have tried to investigate the dynamical behavior of the GP-theories in the framework of the Lagrangian formalism. We have found the explicit form of the Lagrangian constraints for GP theories and showed that they are, in some sense, second class. We have also showed that the Lagrangian approach also verifies that all of GP theories are systems with three dynamical degrees of freedom. 
  
  For this reason we followed a recent Lagrangian approach given in  ref. \cite{a20} following the previous work of \cite{b20}.  There,  a complete A-Z investigation of the Lagrangian approach to constrained systems is proposed. Specially the idea of first class and second class Lagrangian constraints are introduced, and an  accurate formula is given for counting the number of dynamical degrees of freedom. Although, the whole physical description should be equivalent to  Hamiltonian approach, however,  the Lagrangian structure of a system leads to a different points of view for describing and understanding  the physical content of a theory.

This paper is organized as follows. In Section 2 we  review the  approach of ref \cite{a20} in Lagrangian formalism. In section 3, which is the most important  part of the paper, we introduce the GP  theories and investigate their dynamics in the framework of the  Lagrangian formalism of  GP model.  In Section 4 we study  Hamiltonian formalism of different types of the GP  theories. Some detailed calculations  are given in  the Appendix A.


\section{Outline of the analysis}
In this section, we review a certain method proposed in the Lagrangian formalism in order to consider the dynamical behavior of a theory with singular Lagrangian \cite{a20}. 
 The Euler-Lagrange equations of motion for the Lagrangian $L(q, \dot{q}, t)$ with $P$ degrees of freedom are as follows 
 \begin{equation}
L_{i}= W_{ij}\ddot{q}_{j}+\alpha_{i},
\qquad\qquad{i=1,......,P}.\label{01}
 \end{equation}
 where $L_{i}$'s,  are denoted as Eulerian drevatives.  The Hessian matrix elements and $ \alpha_{i}$'s are defined as
\begin{equation}
W_{ij}=\dfrac{\partial^{2 L}}{\partial\dot{q}_{i}\partial\dot{q}_{j}} , \hspace{10mm}
\alpha_{i}=\dfrac{\partial^{2}L}{\partial{q}_{j}\partial\dot{q}_{i}}\dot{q}_{j}-\dfrac{\partial L}{\partial q_{i}}.\label{05}
\end{equation}
 The whole set of accelerations  can  not be determined for a singular Lagrangian.  Suppose  the rank of Hessian matrix  is $(P-A_0)$, leading to $A_0$ null-vectors $\lambda^{a_{0}}$ such that
\begin{equation}
\lambda^{a_{0}}_{i} W_{ij}=0, \hspace{1.2 cm}
a_0=1,\cdots A_0.\label{06}
\end{equation}
 Multiplying both sides of Eq. (\ref{01}) by $\lambda_{i}^{a_0}$ gives 
\begin{equation}
\Gamma^{a_{0}}(q, \dot{q})= \lambda^{a_0}_{i}L_{i} = \lambda^{a_{0}}_{i}\alpha_{i}\approx0, \label{07}
\end{equation} 
where the symbol $\approx$ means weak equality on the  constraint surface,  i.e.  the surface on which the constraints vanish.   Assume for certain combinations $\lambda_{i}^{f_0}$,  of $\lambda_{i}^{a_0}$'s,  the Eqs.  (\ref{07}) are satisfied identically,  leading to $F_0$ Neother identities as 
\begin{equation}
\lambda^{f_{0}}_{i} L_i=0.\label{77}
\end{equation}
The remaining set of null vectors denoted by $\lambda_{i}^{p_0}$'s give $P_0$ first level Lagrangian constraints as $\Gamma^{P_{0}}(q, \dot{q})$.  Then we should  proceed to consistency of first level constraints. To do this we should impose new equations $d\Gamma^{p_{0}}/dt=0$.  In this way we can write
the extended set of equations of motion  as follow
\begin{equation}
W_{i_{1}j}^{1}\ddot{q_{j}}+\alpha_{i_{1}}=0,\qquad\qquad i_{1}=1,\cdots , P, \cdots , P+P_{0},
\end{equation}
where the first $P$ lines of the rectangular matrix $W^1$ is the same as the matrix $W$ and the subsequent lines are in $\dot{\Gamma}^{p_{0}}=0$ for $p_0=1, \cdots , P_0$. 

The extended Hessian  $W^{1}$ may 
 have increased rank $(P-A_0)+S_1$ in comparison with the main Hessian with rank $(P-A_0)$.  This means that we are succeeded to find $S_1$ more accelerations in terms of coordinates and velocities.  The corresponding constraints for this category are denoted as "second class Lagrangian constraints".
 On the other hand,  we should find $A_1=P_0 -S_1$ new null-vectors.  The null vectors $\lambda^{a_1}$ are divided again into $F_1$ null vectors $\lambda^{f_{1}}$ such that $\lambda_{i_{1}}^{f_{1}}L_{i_{1}} \approx0$ (identically) and $P_1$ null vectors $\lambda^{p_{1}}$ where $\Gamma^{p_{1}}\equiv\lambda_{i_{1}}^{p_{1}}L_{i_{1}}$ are the second level Lagrangian constraints.  The new set of Noether identities may be written as
\begin{equation}
\lambda_{i_{1}}^{f_{1}}L_{i_{1}} \equiv \lambda^{f_1}_i L_i+\sum_{p_0=P+1}^{P+P_0}  \lambda^{f_1}_{p_0}\dfrac{d}{dt}(\lambda_j^{p_{0}} L_j)=0, \hspace{15mm} f_1=1, \cdots F_1.\label{n.neo.idn}
\end{equation} 
Eqs.  (\ref{n.neo.idn}) can also be read as 
\begin{equation}
\dfrac{d}{dt} \sum_{p_0=P+1}^{P+P_0}  \lambda^{f_1}_{p_0} \Gamma^{p_{0}} \approx 0
\end{equation} 
where $\approx 0$ means vanishing identically on shell.  In other words we have certain combinations $\Gamma^{f_1} \equiv \sum \lambda^{f_1}_{p_0} \Gamma^{p_{0}}$ which their consistency holds automatically. We call this category of constraints as "first class Lagrangian constraints"\footnote{Do not mix first class and second class with first level,  second level, etc.  At each level we may have first class or second class constraints.}.

In this way $P_0$ first level constraints have been divided into three categories, i.e. $S_1$ second class constraints (which lead to increasing the rank of Hessian), $F_1$ first class constraints (leading to  Noether identities) and $P_1$ pending constraints $\Gamma^{p_0}$ (which lead to the second level constraints $\Gamma^{p_1}$ under consistency), such that $P_0=F_1+P_1+S_1$.

 Investigating consistency of the  second level constraints $\Gamma^{p_{1}}$'s,  gives us similar results as we found at the first level. The process should be performed at subsequent levels of consistency unless there remains no pending constraints at the end of story.
Fig.1 gives a schematic to show how the whole  procedure goes on.
\begin{center}
	\includegraphics[scale=0.6]{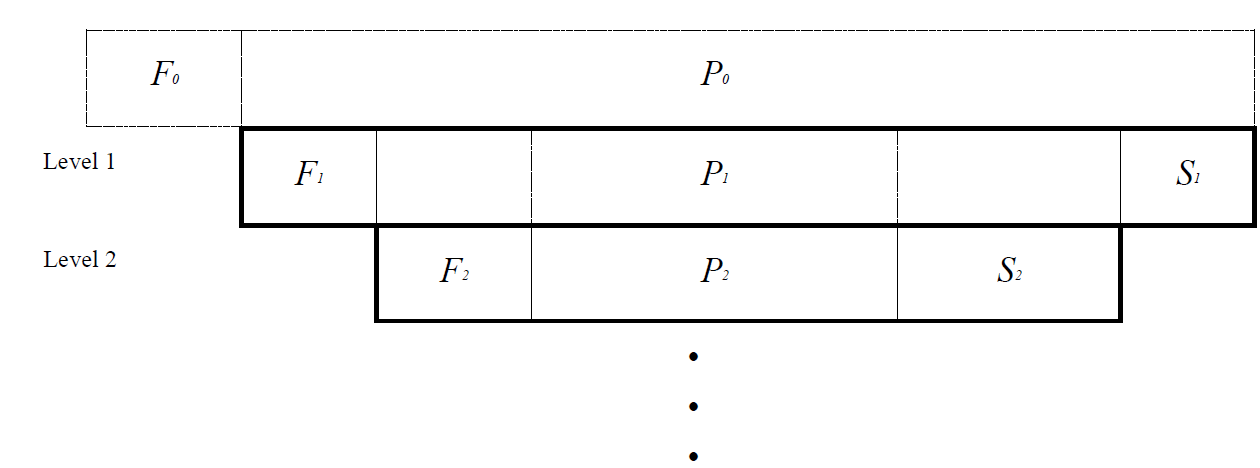}

Fig. 1 -   Diagram of FPS decomposition of Lagrangian constraints
\end{center}
\vspace{3mm}

Regardless of some details, we can manage the whole system of constraints as first class and second class constraint chains. In this "chain structure" every constrain is the result of consistency of its above constraint in the chain. (except for the first level constraints which have been arise from singularity of Hessian.)
When a chain terminates by introducing a Noether identity, the whole set of constraints of the chain are first class. On the other hand,  when a chain terminates by introducing an independent equation for determining the accelerations, then all of the constraints of the corresponding chain are second class.

Our final task in this section is counting the number of dynamical degrees of freedom.  As is shown in details in ref.  \cite{a20}, for a first class constraint chain with $k$ elements we have a gauge transformation including arbitrary gauge parameter $\eta(t)$ and its time derivatives up to $d^k \eta / dt^k$,  i.e. $k+1$ gauge parameters altogether. Hence, if $F$ denotes the total number of Noether identities, and $FC$ denotes the total number of first class Lagrangian constraints, the number of dynamical variables is decreased by $F+FC$ according to gauge symmetries. On the other hand, each second class Lagrangian constraint imposes just one limitation on the choice of initial conditions in the space of coordinates and velocities, while each dynamical degree of freedom corresponds to two initial conditions.  Therefore, If $SC$ denotes the total number of second class constraints, our final formula for counting the number of dynamical degrees of freedom reads
 \begin{equation}
\mathcal{N}=P-(F.C+F)-\frac{1}{2}S.C. \label{a20}
\end{equation}

  \section{generalized Proca theory }
 The Proca action is given by
\begin{equation}
\mathcal S_{\rm Proca}  = \int \mathrm{d}^4x \left[-\frac14 F_{\mu\nu}^2 -\frac12 m^2 A^2 \right],  \label{ProcaAction} 
\end{equation}
where $F_{\mu\nu}=\partial_\mu A_\nu -\partial_\nu A_\mu$.
The Lagrangian of the generalized Proca model (GP) \cite{a14} is achieved by adding certain self-interactions, including the derivatives of the vector field, as follows
\begin{equation}
\mathcal L_{\rm gen. Proca} = -\frac14 F_{\mu\nu}^2 +\sum^5_{n=2}\alpha_n \mathcal L_n, \label{eq1}
\end{equation}
in which  the self-interactions of the vector field have the following forms 
\begin{eqnarray}
\mathcal L_2 & = &f_2(A_\mu, F_{\mu\nu}, \tilde{F}_{\mu\nu}),\nonumber\\
\mathcal L_3 & = &f_3(A^2) \;\; \partial\cdot A ,\\
\mathcal L_4  &=&  f _4(A^2)\;\left[(\partial\cdot A)^2-\partial_\rho A_\sigma \partial^\sigma A^\rho\right] + c_2\tilde{f} _4(A^2)F^2 ,\nonumber\\
\mathcal L_5  &=&f_5(A^2)\;\left[(\partial\cdot A)^3-3(\partial\cdot A)\partial_\rho A_\sigma \partial^\sigma A^\rho +2\partial_\rho A_\sigma \partial^\gamma A^\rho\partial^\sigma A_\gamma  \right]  \nonumber\\
&& 
+d_2\tilde{f}_5(A^2)\tilde{F}^{\mu\alpha}\tilde{F}_{\alpha}^\nu\partial_\mu A_\nu , \nonumber\\
\mathcal L_6  &=&e_2f_6(A^2) \tilde{F}^{\alpha\beta}\tilde{F}^{\mu\nu}\partial_\alpha A_\mu \partial_\beta A_\nu , \label{veq}\nonumber
\end{eqnarray}
and  $\partial\cdot A=\partial_\mu A^\mu$.  The functions $f_{2,3,4,5}$ are arbitrary functions of  $A^2=A_\mu A^\mu$,  however,  the function $f_2$ may also depend on $F_{\mu\nu}$ and $\tilde{F}_{\mu\nu}$ (where $\tilde{F}$ is the dual of F).  The above terms are derived by considering all possible contractions of the tensor $\epsilon^{\mu\nu\alpha\beta}$ with the derivative tensor $\partial_\mu A_\nu$ as follows
\begin{eqnarray}\label{epsi1}
\mathcal L_2&=&-\frac{f_2(A_\mu, F_{\mu\nu}, \tilde{F}_{\mu\nu})}{24}\mathcal{E}^{\mu\nu\alpha\beta}\mathcal{E}_{\mu\nu\alpha\beta} =f_2(A_\mu, F_{\mu\nu}, \tilde{F}_{\mu\nu}) \nonumber\\
\mathcal L_3&=&-\frac{f_3(A^2)}{6}\mathcal{E}^{\mu\nu\alpha\beta}\mathcal{E}^{\rho}_{\;\;\;\nu\alpha\beta} \partial_\mu A_\rho=f_3(A^2)\;\; \partial\cdot A \nonumber\\
\mathcal{L}_4&=&-\frac{1}{2}\mathcal{E}^{\mu\nu\rho\sigma}\mathcal{E}^{\alpha\beta}_{\;\;\;\;\rho\sigma}(f_4(A^2)\partial_\mu A_\alpha\partial_\nu A_\beta+c_2\tilde{f}_4(A^2)\partial_\mu A_\nu\partial_\alpha A_\beta)\nonumber\\
&=&f _4(A^2)\; \left[(\partial\cdot A)^2-\partial_\rho A_\sigma \partial^\sigma A^\rho\right]+c_2 \tilde{f} _4(A^2)F_{\rho\sigma}^2 \\
\mathcal{L}_5&=&-\mathcal{E}^{\mu\nu\rho\sigma}\mathcal{E}^{\alpha\beta\delta}_{\;\;\;\;\;\;\sigma}(f_5(A^2)\partial_\mu A_\alpha\partial_\nu A_\beta \partial_\rho A_\delta +d_2\tilde{f}_5(A^2)\partial_\mu A_\nu\partial_\rho A_\alpha \partial_\beta A_\delta) \nonumber\\
& =&f_5(A^2)\left[(\partial\cdot A)^3-3(\partial\cdot A)\partial_\rho A_\sigma \partial^\sigma A^\rho 
+2\partial_\rho A_\sigma \partial^\gamma A^\rho\partial^\sigma A_\gamma \right] \nonumber\\
&+&d_2\tilde{f}_5(A^2)\tilde{F}^{\alpha\mu}\tilde{F}^\beta_{\;\;\mu}\partial_\alpha A_\beta \nonumber\\
\mathcal{L}_6&=&-\mathcal{E}^{\mu\nu\rho\sigma}\mathcal{E}^{\alpha\beta\delta\kappa}(f_6(A^2)\partial_\mu A_\alpha\partial_\nu A_\beta \partial_\rho A_\delta \partial_\sigma A_\kappa +e_2\tilde{f}_6(A^2)\partial_\mu A_\nu\partial_\alpha A_\beta \partial_\rho A_\delta  \partial_\sigma A_\kappa)\nonumber\\
&=&f_6(A^2)\left[3\partial^\beta A^\alpha(\partial_\alpha A_\beta \partial_\mu A_\nu \partial^\nu A^\mu-2\partial_\alpha A_\nu \partial^\mu A_\beta \partial^\nu A_\mu)+8(\partial\cdot A)\partial_\beta A_\nu \partial^\mu A^\beta \partial^\nu A_\mu \right. \nonumber\\
&&\left.-6 (\partial\cdot A)^2\partial_\mu A_\nu \partial^\nu A_\mu+(\partial\cdot A)^4 \right] \nonumber\\
&+&e_2\tilde{f}_6(A^2)\Big[ (\partial\cdot A)(2\partial^\gamma A^\beta(\partial_\beta A_\delta \partial^\delta A_\gamma-\partial_\delta A_\gamma \partial^\delta A_\beta)+(\partial\cdot A)(\partial_\delta A_\gamma - \partial_\gamma A_\delta)\partial^\delta A^\gamma)) \nonumber\\
&+&\partial^\beta A^\alpha(-2\partial_\alpha A_\delta \partial^\gamma A_\beta \partial^\delta A_\beta+\partial_\gamma A_\delta(2\partial^\gamma A_\alpha \partial^\delta A_\beta+(\partial_\alpha A_\beta -\partial_\beta A_\alpha)\partial^\delta A^\gamma))\Big]\nonumber\\
&=&e_2\tilde{f}_6(A) \tilde{F}^{\alpha\beta}\tilde{F}^{\mu\nu}\partial_\alpha A_\mu \partial_\beta A_\nu .\nonumber
\end{eqnarray}
 It is worth noting that the sixth order Lagrangian,
as it has been explicitly shown in \cite{a18},  is  just a total derivative in four-dimensional flat space-time and so can be ignored.


We will consider the special case where the arbitrary functions are considered  as  $f_{2,3,4,5}=A^2$. Thus the Lagrangian terms in (\ref{epsi1}) reads as follow
\begin{eqnarray}
\mathcal L_2 & = &A^2,\nonumber\\
\mathcal L_3 & = &A^2(\partial\cdot A), \nonumber\\
\mathcal L_4  &=& A^2\left[(\partial\cdot A)^2+c_2\partial_\rho A_\sigma \partial^\rho A^\sigma-(1+c_2)\partial_\rho A_\sigma \partial^\sigma A^\rho\right],   \nonumber\\
\mathcal L_5  &=& A^2\left[(\partial\cdot A)^3-3d_2(\partial\cdot A)\partial_\rho A_\sigma \partial^\rho A^\sigma \right. \left.
-3(1-d_2)(\partial\cdot A)\partial_\rho A_\sigma \partial^\sigma A^\rho \right. \\
&&\left. +2\left(1-\frac{3d_2}{2}\right)\partial_\rho A_\sigma \partial^\gamma A^\rho\partial^\sigma A_\gamma 
 \right. \left. +2\left(\frac{3d_2}{2}\right)\partial_\rho A_\sigma \partial^\gamma A^\rho\partial_\gamma A^\sigma\right]  \nonumber.
\end{eqnarray}
One main point is that, due to constructions given in Eqs.   (2) and (4),  all of the terms in GP models do not involve  time derivatives of $A_0$.  Hence,  we should be prepared to investigate a constrained system either in Lagrangian or in the Hamiltonian formalism.

\section{Lagrangian formalism of the GP for especially case}\label{specialcase}
The equations of motion for the Lagrangian given by Eq.  (\ref{eq1}) is as follows
 \begin{eqnarray}
 -\partial_\nu(\partial^\mu A^\nu-\partial^\nu A^\mu)+\sum_{i=2}^{5} \alpha_i \mathcal E_{i}=0,
\end{eqnarray}
where

\begin{eqnarray}
\mathcal E_{2} & = & 2A_\mu,\nonumber\\
\mathcal E_3 & = &2A_\mu(\partial\cdot A) -2A^\nu \partial_\mu A_\nu, \nonumber\\
\mathcal E_4 & = & 2\Big(A_\mu \left[(\partial\cdot A)^2-(1+c_2)\partial_\rho A_\sigma \partial^\sigma A^\rho+c_2\partial_\rho A_\sigma \partial^\sigma A^\rho\right]
+c_2 A^2(-\Box A_\mu+\partial_\nu\partial_\mu A^\nu)
\nonumber\\
&&-2c_2 A^\rho \partial_\nu A_\rho \partial^\nu A_\mu-2(\partial\cdot A) A^\rho \partial_\mu A_\rho +2(1+c_2)A^\rho \partial_\nu A_\rho \partial_\mu A^\nu \Big) , \nonumber\\
\mathcal E_5 & = &2A_\mu\left[(\partial\cdot A)^3+3(-1+d_2)(\partial\cdot A)\partial_\rho A_\sigma \partial^\sigma A^\rho \right. \\
&& \left.-3d_2(\partial\cdot A)\partial_\rho A_\sigma \partial^\rho A^\sigma+(2-3d_2)\partial_\rho A_\sigma \partial^\gamma A^\rho\partial^\sigma A_\gamma \right.  \left. +3d_2\partial_\rho A^\sigma \partial^\rho A^\gamma\partial_\sigma A_\gamma\right]
 -3A^\rho\Big( -d_2(4\partial_\nu A_\rho \partial^\nu A_\mu (\partial\cdot A) \nonumber\\
&& -2(\partial_\nu A_\mu \partial^\nu A_\sigma +\partial^\nu A_\mu \partial_\sigma A_\nu)\partial^\sigma A_\rho  +A_\rho(\partial^\nu A_\mu(\partial_\sigma\partial_\nu A^\sigma -\Box A_\nu)+2(\partial\cdot A)(\Box A_\mu - \partial_\sigma\partial_\mu A^\sigma)  \nonumber\\
&&+(\partial_\nu\partial_\mu A_\sigma-2\partial_\sigma\partial_\nu A_\mu+\partial_\sigma\partial_\mu A_\nu)\partial^\sigma A^\nu))   \nonumber\\
&&+2((\partial\cdot A)^2+((-1+d_2)\partial_\nu A_\sigma -d_2\partial_\sigma A_\nu)\partial^\sigma A^\nu)\partial_\mu A_\rho  +(4(-1+d_2)\partial_\nu A_\rho (\partial\cdot A) \nonumber\\
&&+d_2A_\rho(-\partial_\sigma\partial_\nu A^\sigma+\Box A_\nu)  +2((2-3d_2)\partial_\nu A_\sigma + d_2 \partial_\sigma A_\nu)\partial^\sigma A_\rho)\partial_\mu A^\nu\Big).\nonumber
\end{eqnarray}
In the following we consider different terms one by one.
 \subsection{Lagrangian formalism for $\mathcal{L}_2$ }
Consider the original Proca model consisting of  the Maxwell kinetic term besides the  $\mathcal L_2$ term.  In this case    the equations of motion read
\begin{eqnarray}
   -\partial_\nu(\partial^\mu A^\nu-\partial^\nu A^\mu)+2A_\mu=0,
\end{eqnarray}
Let us separate different components of the vector field and write the equations of motion in the matrix form as
\begin{eqnarray}\label{veq}
\begin{pmatrix}
0&0&0&0 \\
0&1 &0&0 \\
0&0&1 &0 \\
0&0&0&1
 \end{pmatrix}
 \begin{pmatrix}
\partial_0 \partial^0 A^0  \\
\partial_0 \partial^0 A^1 \\
\partial_0 \partial^0 A^2 \\
\partial_0 \partial^0 A^3
 \end{pmatrix}+
  \begin{pmatrix}
\partial_i \partial^i A^0- \partial_i \partial^0 A^i+2A^0\\
-\partial_1 \partial^0 A^0+ \partial_i \partial^i A^1-\partial_i \partial^1 A^i+2A^1\\
-\partial_2 \partial^0 A^0+ \partial_i \partial^i A^2-\partial_i \partial^2 A^i+2A^2\\
-\partial_3 \partial^0 A^0+ \partial_i \partial^i A^3-\partial_i \partial^3 A^i+2A^3
 \end{pmatrix}=0.\label{EH}
\end{eqnarray}
Following the algorithm given in \cite{a20} and \cite{b20},  the null vector of the above Hessian matrix has the following form
\begin{eqnarray}
\lambda_{(0)}=(1,0,0,0),\label{nv}
\end{eqnarray}
Multiplying both sides of Eq. (\ref{EH}),  from the left,   by the null vector (\ref{nv}),  we find the first level Lagrangian  constraint as 
\begin{eqnarray}
C_1=\partial_i \partial^i A^0- \partial_i \partial^0 A^i+2A^0 =0,
\end{eqnarray}
Then we should annex a new row to the matrices of the Eq.  (\ref{EH}) by differentiating the first level constraint with respect to time.  This guarantees consistency of the constraint in the course of the time.  Hence,  we have  
\begin{eqnarray}\label{cc}
\begin{pmatrix}
0&0&0&0 \\
0&1 &0&0 \\
0&0&1 &0 \\
0&0&0&1\\
0&-\partial_1&-\partial_2&-\partial_3
 \end{pmatrix}
 \begin{pmatrix}
\partial_0 \partial^0 A^0  \\
\partial_0 \partial^0 A^1 \\
\partial_0 \partial^0 A^2 \\
\partial_0 \partial^0 A^3
 \end{pmatrix}+
  \begin{pmatrix}
\partial_i \partial^i A^0- \partial_i \partial^0 A^i+2A^0\\
-\partial_1 \partial^0 A^0+ \partial_i \partial^i A^1-\partial_i \partial^1 A^i+2A^1\\
-\partial_2 \partial^0 A^0+ \partial_i \partial^i A^2-\partial_i \partial^2 A^i+2A^2\\
-\partial_3 \partial^0 A^0+ \partial_i \partial^i A^3-\partial_i \partial^3 A^i+2A^3\\
\partial_i \partial^i \partial_0 A^0+2\partial_0 A^0
 \end{pmatrix}=0 .
\end{eqnarray}
The extended Hessian is still a rank-3 matrix.  In other words, we have a new (operator valued) null-vector as
\begin{eqnarray}
\lambda_{(1)}=(0,\partial_1,\partial_2,\partial_3,1) .
\end{eqnarray}
Again,  multiplying by $\lambda_{(1)}$ from the left,  Eq. (\ref{cc}) gives the second level constraint  in the following form
 \begin{eqnarray}
C_2=\partial_\mu A^\mu=0.
\end{eqnarray}
Consistency of this new constraint requires considering the time derivative of this constraint as the next additional row of the matrices of Eq.  (\ref{cc}),  as follows
\begin{eqnarray}
\begin{pmatrix}
0&0&0&0 \\
0&1 &0&0 \\
0&0&1 &0 \\
0&0&0&1\\
0&-\partial_1&-\partial_2&-\partial_3\\
2&0&0&0 
 \end{pmatrix}
 \begin{pmatrix}
\partial_0 \partial^0 A^0  \\
\partial_0 \partial^0 A^1 \\
\partial_0 \partial^0 A^2 \\
\partial_0 \partial^0 A^3
 \end{pmatrix}+
  \begin{pmatrix}
\partial_i \partial^i A^0- \partial_i \partial^0 A^i+2A^0\\
-\partial_1 \partial^0 A^0+ \partial_i \partial^i A^1-\partial_i \partial^1 A^i+2A^1\\
-\partial_2 \partial^0 A^0+ \partial_i \partial^i A^2-\partial_i \partial^2 A^i+2A^2\\
-\partial_3 \partial^0 A^0+ \partial_i \partial^i A^3-\partial_i \partial^3 A^i+2A^3\\
\partial_i \partial^i \partial_0 A^0+2\partial_0 A^0\\
2\partial_i \partial_0 A^i
 \end{pmatrix}=0.
\end{eqnarray}
At this level,  we have no new null vector and consequently no new constraints.  In fact,  the $4 \times 6$ extended Hessian matrix is maximal rank, i.e. rank-4.  Thus, the two constraints ($C_1$ and $C_2$) are second class Lagrangian constraints (in the terminology of ref.  \cite{a20}).  In this way we have two second class and no first class constraint.  Using the master Eq. (\ref{a20}) for counting degrees of freedom we have
\begin{eqnarray}
D OF= \mathcal {N}-(F.C+F)-\frac{1}{2}S.C=4-\frac{1}{2}(2)=3,
\end{eqnarray}
where $\mathcal{N}$ is total number of degrees of freedom, F.C is the number of  first class constraints,  S.C is the number of second class constraints and $F$ is the number of gauge parameters.  Hence,  in Proca model with $\mathcal{L}_2$ term, we have three dynamical fields and no gauge variable.
 \subsection{Lagrangian formalism for $\mathcal{L}_3$,$\mathcal{L}_4$ and $\mathcal{L}_5$ }
In this subsection we consider the equations of motion for the Lagrangian $\mathcal{L}_3$ added to the Maxwell kinetic term.  The equations of motion read as follows
\begin{eqnarray}
 -\partial_\nu(\partial^\mu A^\nu-\partial^\nu A^\mu)+2A_{\mu}\partial_\nu A^\nu-2A^\nu\partial_\mu A_\nu=0.
\end{eqnarray}
Separating different components of the fields, the matrix form of the equations of motion read
\begin{eqnarray}
\begin{pmatrix}
0&0&0&0 \\
0&1 &0&0 \\
0&0&1 &0 \\
0&0&0&1
 \end{pmatrix}
 \begin{pmatrix}
\partial_0 \partial^0 A^0  \\
\partial_0 \partial^0 A^1 \\
\partial_0 \partial^0 A^2 \\
\partial_0 \partial^0 A^3
 \end{pmatrix}+
  \begin{pmatrix}
\partial_i \partial^i A^0- \partial_i \partial^0 A^i+2A_{0}\partial_i A^i-2A^i \partial_0 A_i
\\
-\partial_1 \partial^0 A^0+ \partial_i \partial^i A^1-\partial_i \partial^1 A^i+2A_{1}\partial_0 A^0-2A^0\partial_1 A_0+2A_{1}\partial_i A^i-2A^i\partial_1 A_i
\\
-\partial_2 \partial^0 A^0+ \partial_i \partial^i A^2-\partial_i \partial^2 A^i+2A_{2}\partial_0 A^0-2A^0\partial_2 A_0+2A_{2}\partial_i A^i-2A^i\partial_2 A_i
\\
-\partial_3 \partial^0 A^0+ \partial_i \partial^i A^3-\partial_i \partial^3 A^i+2A_{3}\partial_0 A^0-2A^0\partial_3 A_0+2A_{3}\partial_i A^i-2A^i\partial_3 A_i
 \end{pmatrix}=0. \label{a25}
\end{eqnarray}
The null vector of the  Hessian matrix has the following form
\begin{eqnarray}
\lambda_{(0)}=(1,0,0,0).
\end{eqnarray}
Multiplying  the equation of motion,  from the left, by  $\lambda_{(0)}$   we have one first level constraint as 
 \begin{eqnarray}
C_1= \partial_i \partial^i A^0- \partial_i \partial^0 A^i+2A^0+2A_{0}\partial_i A^i-2A^i \partial_0 A_i=0.
\end{eqnarray}
in the next step,  we should consider the time derivative of the  constraint  $C_1$ together with the equation of motion,  as follows
\begin{eqnarray}&&
\begin{pmatrix}
0&0&0&0 \\
0&1 &0&0 \\
0&0&1 &0 \\
0&0&0&1\\
0&-\partial_1-2A_1&-\partial_2-2A_2&-\partial_3-2A_3
 \end{pmatrix}
 \begin{pmatrix}
\partial_0 \partial^0 A^0  \\
\partial_0 \partial^0 A^1 \\
\partial_0 \partial^0 A^2 \\
\partial_0 \partial^0 A^3
 \end{pmatrix}+
  \nonumber
 \\&&
  \begin{pmatrix}
\partial_i \partial^i A^0- \partial_i \partial^0 A^i+2A_{0}\partial_i A^i-2A^i \partial_0 A_i
\\
-\partial_1 \partial^0 A^0+ \partial_i \partial^i A^1-\partial_i \partial^1 A^i+2A_{1}\partial_0 A^0-2A^0\partial_1 A_0+2A_{1}\partial_i A^i-2A^i\partial_1 A_i
\\
-\partial_2 \partial^0 A^0+ \partial_i \partial^i A^2-\partial_i \partial^2 A^i+2A_{2}\partial_0 A^0-2A^0\partial_2 A_0+2A_{2}\partial_i A^i-2A^i\partial_2 A_i
\\
-\partial_3 \partial^0 A^0+ \partial_i \partial^i A^3-\partial_i \partial^3 A^i+2A_{3}\partial_0 A^0-2A^0\partial_3 A_0
+2A_{3}\partial_i A^i-2A^i\partial_3 A_i
\\
\partial_0(\partial_i \partial^i A^0+2 A^0  \partial_i A^i)-2 \partial_0 A^i \partial^0 A_i
 \end{pmatrix}=0.\label{extendedequation3}
\end{eqnarray}
We can find a new null vector for  the  above Hessian matrix as 
\begin{eqnarray}
\lambda_{(1)}=(0,\partial_1+2A_1,\partial_2+2A_2,\partial_3+2A_3,1).  \label{nv2l3}
\end{eqnarray}
Multiplying the extended equations (\ref{extendedequation3}) by the new null vector (\ref{nv2l3}) gives the second level Lagrangian 
 constraint in the following form 
\begin{eqnarray}
C_2=(\partial_j+2A_j)(-\partial^j \partial_0 A^0+\partial_i \partial^iA^j-\partial_i \partial^jA^i+2A_j\partial_0A^0-2A^0\partial_jA_0+2A_j\partial_i A^i-2A^i\partial_jA_i)=0,
\end{eqnarray}
Then we should annex the time derivative of the  constraint $C_2$ to the extended equations (\ref{extendedequation3}),  to get the following
\begin{eqnarray}
\begin{pmatrix}
0&0&0&0 \\
0&1 &0&0 \\
0&0&1 &0 \\
0&0&0&1\\
0&-\partial_1&-\partial_2&-\partial_3\\
2+(\partial_j+2A_j)(-\partial^j+2A_j)&0&0&0 
 \end{pmatrix}
 \begin{pmatrix}
\partial_0 \partial^0 A^0  \\
\partial_0 \partial^0 A^1 \\
\partial_0 \partial^0 A^2 \\
\partial_0 \partial^0 A^3
 \end{pmatrix}+
  \begin{pmatrix}
\partial_i \partial^i A^0- \partial_i \partial^0 A^i+..\\
-\partial_1 \partial^0 A^0+ \partial_i \partial^i A^1-\partial_i \partial^1 A^i+...\\
-\partial_2 \partial^0 A^0+ \partial_i \partial^i A^2-\partial_i \partial^2 A^i+...\\
-\partial_3 \partial^0 A^0+ \partial_i \partial^i A^3-\partial_i \partial^3 A^i+...\\
\partial_i \partial^i \partial_0 A^0+2\partial_0 A^0+...\\
2\partial_i \partial_0 A^i+...
 \end{pmatrix}=0
\end{eqnarray}
Again the final $4 \times 6$ Hessian matrix is full rank.  So,  we do not have any new constraint and the two Lagrangian constraints $C_1$ and $C_2$ are second class constraints.  Counting the number of degrees of freedom according to the Eq. (\ref{a20}) gives
\begin{eqnarray}
DOF= \mathcal {N}-(F.C+F)-\frac{1}{2}S.C=4-\frac{1}{2}(2)=3.
\end{eqnarray}
The same procedure can be followed for $\mathcal{L}_4$ and $\mathcal{L}_5$. The main point is that the equations of motion do not contain the term $\partial^0 \partial_0 A^0$. So, similar to Eqs (\ref{EH}) and (\ref{a25}) the Hessian has the following form 
\begin{eqnarray}
H_{\mu \nu}=
\begin{pmatrix}
0&0&0&0 \\
0&1 &0&0 \\
0&0&1 &0 \\
0&0&0&1
 \end{pmatrix}.
\end{eqnarray}
As before, $\lambda_{(0)}=(1,0,0,0)$
is null-vector. For $\mathcal{L}_4$ the primary constraint emerges as 
\begin{eqnarray}
C_1 & = & 2\Big(A_0 \left[(\partial\cdot A)^2-(1+c_2)\partial_\rho A_\sigma \partial^\sigma A^\rho+c_2\partial_\rho A_\sigma \partial^\sigma A^\rho\right]+c_2 A^2(-\Box A_0+\partial_\nu\partial_0 A^\nu)
\nonumber\\
&&-2c_2 A^\rho \partial_\nu A_\rho \partial^\nu A_0-2(\partial\cdot A) A^\rho \partial_0 A_\rho +2(1+c_2)A^\rho \partial_\nu A_\rho \partial_0 A^\nu \Big) ,
\end{eqnarray}
for $ \mathcal{L}_5$ a more involved primary constraint emerges, as well.

For both cases the primary constraint   does not contain $ \partial_0 A^0$. Hence, the extended 
 Hessian appears as 
\begin{eqnarray}
H_{\mu \nu}=
\begin{pmatrix}
0&0&0&0 \\
0&1 &0&0 \\
0&0&1 &0 \\
0&0&0&1\\
0&\mathcal{X}&\mathcal{Y}&\mathcal{Z}
 \end{pmatrix}
\end{eqnarray}
where $\mathcal{X}$, $\mathcal{Y}$ and $\mathcal{Z}$   are functions of  $ A^\mu$ and  $ \partial_\mu A^\mu$. The extended Hessian $H_{\mu\nu}$
is not full rank yet. Thus, we have another null vector as follows 
\begin{eqnarray}
\lambda_{(1)}=(0,-\mathcal{X},-\mathcal{Y},-\mathcal{Z},1).
\end{eqnarray}
Multiplying the extended equations of motion with $\lambda_{(1)}$ gives the secondary constraint $C_2$
in each case. The detailed form of $C_2$ does not bring any important information except that $C_2$ includes the term  $  \partial_0 A^0$. This leads to increasing the rank of the extended Hessian at the next level. Hence, we have no gauge symmetry, and two second class constraints $C_1$ and $C_2$.
The number of degrees of freedom comes out to be 3 as in the previous cases for $\mathcal{L}_2$ and $\mathcal{L}_3$.  

\section{Hamiltonian formalism }
In this section, we consider the Hamiltonian structure of the Proca theory. As we will see the results differ from Maxwell theory. The Lagrangian of Proca theory has the following form 
\begin{eqnarray}
S=\int d^4x (-\frac{1}{4}F^{\mu\nu}F_{\mu\nu}+m^2A^{\mu}A_\mu),  
\end{eqnarray}
where  $m^2$ is the mass parameter. The absence of  $\partial_0 A^0$ leads to the following  primary constraint 
\begin{eqnarray}
C_1\equiv \Pi^0 \approx 0.
\end{eqnarray}
The momentum fields conjugate to $A^i$ read
\begin{eqnarray}
\Pi^i=\partial_0A^i-\partial^{i}A_{0}.\label{momen}
\end{eqnarray}
Thus, the total Hamiltonian reads
\begin{eqnarray}
H_{t} \equiv \int d^3x( \mathcal{H}_c+\lambda C_1),
\end{eqnarray}
where $\lambda$ is  undetermined Lagrange multiplier and
\begin{eqnarray}
 \mathcal{H}_c= \frac{1}{2}\Pi^i\Pi_i+\frac{1}{2}F^{ij}F_{ij}-A_{0}\partial_{i}\Pi^{i}-m^2((A^0)^2+A^iA_i).
\end{eqnarray}
Consistency of the primary constraint gives
\begin{eqnarray}
C_2(x)\equiv \left\{ C_1(x),H_t\right\}=\partial_{i}\Pi^i(x)+2m^2A^0(x).
\end{eqnarray}
Consistency of the secondary constraint  gives
\begin{eqnarray}
0 \approx \{C_2(x), H_t\}=2m^2\partial_iA^i+2m^2\lambda. 
\end{eqnarray}
Hence the Lagrange multiplayer is determined as $\lambda=-\partial_i A^i$. This means that the constraints  $C_1$ and $C_2$ are second class. 
Using the master formula in Hamiltonian formalism 
\begin{eqnarray}
DoF=\mathcal N -2 \times F.C-S.C,
\end{eqnarray}
in which $\mathcal N$ is total number of phase space variables, F.C is the number of  first class constraints,  S.C is the number of second class constraints.
Counting degrees of freedom gives 
\begin{eqnarray}
DoF=8-2=6, 
\end{eqnarray}
which corresponds to 3 degrees of freedom in configuration space. 

For GP theory with $\mathcal L_3$ we have 
\begin{eqnarray}
S=\int d^4x (-\frac{1}{4}F^{\mu\nu}F_{\mu\nu}+m^2A^{\mu}A_\mu(\partial_\nu A^\nu)),  
\end{eqnarray}
the momenta are similar to Eq. (\ref{momen}) and the primary constraint is $C_1=\Pi^0-A^2$.
The canonical Hamiltonian reads 
\begin{eqnarray}
\mathcal {H}_c= \frac{1}{2}\Pi^i\Pi_i+\frac{1}{2}F^{ij}F_{ij}-A_{0}\partial_{i}\Pi^{i}+m^2((A^0)^2+A^iA_i)\partial_jA^j.
\end{eqnarray}
The secondary constraint in this case is 
\begin{eqnarray}
C_2(x)=\partial_{i}\Pi^i-2A^0m^2\partial_jA^j-2A^j\Pi_j-2\partial_i(A_0A^i).
\end{eqnarray}
Consistency of $C_2(x)$ leads to 
\begin{eqnarray}
0\approx \{C_2(x),\int d^3y(\mathcal{H}_{c}(y)+\lambda C_1(y))\}=\{C_2(x),\int d^3yH_c(y)\}-\lambda(4A_jA^j+(2m^2-4)\partial_jA^j),
\end{eqnarray}
which  determined  $\lambda$. Again, $C_1$ and $C_2$ are second class and we have 6 degrees of freedom in phase space. 

Similar procedure can be followed simply for  $\mathcal L_4$ and  $\mathcal L_5$ . In each case we have two second class constraints and the number of degrees of freedom came out to be 3 in configuration space.  This shows complete agreement with our results in the Lagrangian formalism.

In summary, in this work we explained a new Lagrangian approach \cite{a20} for analyzing the dynamical content of GP theories. For each type of the models we found a Lagrangian constraint chain with two elements. These are second class in a Lagrangian interpretation. The master formula for counting the Lagrangian degrees of freedom (i.e. Eq. (\ref{a20})) shows that all types of GP theories possess 3 dynamical degrees of freedom without any gauge invariance. In order to complete our study, we also followed the traditional Dirac procedure to find the Hamiltonian constraint structure of the system. This was not done in the present literature.


\appendix


  \section{ Hamiltonian formalism of Maxwell theory \label{app:Hamilton}}
 Let us reviewing  Hamiltonian formalism of Maxwell theory
\begin{eqnarray}
S=\int d^4x -\frac{1}{4}F^{\mu\nu}F_{\mu\nu},  
\end{eqnarray}
where 
$F_{\mu\nu}=\partial_{\mu}A_{\nu}-\partial_{\nu}A_{\mu}.
$  In this case, the absence of  $\partial_0 A^0$ leads to the following  primary constraint
\begin{eqnarray}
C_1\equiv \Pi^0\approx0,
\end{eqnarray}
the momentum fields conjugate to $A^i$ reads
\begin{eqnarray}
\Pi^i=\partial_0A^i-\partial^{i}A_{0},
\end{eqnarray}
Hence, the total Hamiltonian is 
\begin{eqnarray}
H_{t}=\int d^3x ( \mathcal{ H}_c+\lambda C_1),
\end{eqnarray}
where $\lambda$ is  undetermined Lagrange multiplier and 
\begin{eqnarray}
\mathcal{ H}_c=\frac{1}{2}\Pi^i\Pi_i+\frac{1}{2}F^{ij}F_{ij}-A_{0}\partial_{i}\Pi^{i}.
\end{eqnarray}
Consistency of the primary constraint leads to the following secondary constraint 
\begin{eqnarray}
C_2(x)\equiv \{\Pi^0(x),\int d^3yH_{c}(y)+\lambda C_1(y)\}=\partial_{i}\Pi^i \approx 0.
\end{eqnarray}
Thus, considering consistency of the secondary constraint $C_2$, we have 
\begin{eqnarray}
\dot{C_2}(x)=\{\partial_{k} \Pi^k(x),H_{t}\}=0  .
\end{eqnarray}
thus we find all together two first class constraints.
 Counting the number of degrees of freedom is
\begin{eqnarray}
DoF=8-4=4, 
\end{eqnarray}
which corresponds to 2 degrees of freedom in configuration space.


\bibliographystyle{}
\bibliography{main}

\end{document}